\documentclass[12pt,reqno]{article}
\usepackage{amsfonts}
\usepackage{amsmath}
\usepackage{amsbsy}

\usepackage{amssymb,latexsym}
\numberwithin{equation}{section}

\begin{document}
 \allowdisplaybreaks[1]
\title{Symmetric Space $\sigma$-model Dynamics: Current Formalism}
\author{Nejat T. Y$\i$lmaz\\
Department of Mathematics
and Computer Science,\\
\c{C}ankaya University,\\
\"{O}\u{g}retmenler Cad. No:14,\quad  06530,\\
 Balgat, Ankara, Turkey.\\
          \texttt{ntyilmaz@cankaya.edu.tr}}
\maketitle
\begin{abstract}
After explicitly constructing the symmetric space sigma model
lagrangian in terms of the coset scalars of the solvable Lie
algebra gauge in the current formalism we derive the field
equations of the theory.
\end{abstract}

\section{Introduction}
The low energy effective limit or the massless background coupling
of the superstring theories are governed by the supergravity
theories \cite{kiritsis}. The scalar sectors of the supergravity
theories reflect the global symmetry properties of these theories
\cite{westbook,westsugra,tani}. Also the restriction of the global
symmetry group to the integers gives the U-duality symmetry of the
corresponding superstring theory \cite{nej125,nej126}. The
majority of the scalar sectors of the supergravities can be
constructed as $G/K$ symmetric space sigma models
\cite{westsugra,tani,julia1,julia2,ker1,ker2}. The symmetric space
sigma model field equations in the internal metric formalism
\cite{julia1,julia2} have been studied in \cite{ker1,ker2} and
further in \cite{nej1,nej2}. In these works the field equations
are obtained under a specific trace convention of the symmetry
algebra representation. In \cite{sssm1} in the internal metric
formalism, for a generic trace convention the most general form of
the field equations are obtained for the axion-dilaton
parametrization of the symmetric space coset manifold. On the
other hand in the vielbein formalism
\cite{westbook,westsugra,tani} without specifying a gauge which
would parameterize the symmetric space coset manifold the
derivation of the field equations is a standard task which results
in a set of equations for the vielbein which is on the symmetric
space $G/K$.

In this work we assume the solvable Lie algebra gauge \cite{fre}
to parameterize the scalar coset manifold $G/K$. This
parametrization contributes a simplification to the construction
of the lagrangian in the current formalism which in this case can
exactly be expressed in terms of the Cartan-form instead of only
an abstractly defined piece of it \cite{westbook,westsugra,tani}.
Thus by using the Cartan-form which is derived in \cite{nej2} we
will explicitly construct the lagrangian in terms of the scalar
fields which parameterize the symmetric space $G/K$. We will then
derive the field equations of the theory for the coset scalars by
directly varying the lagrangian which we have constructed in the
solvable Lie algebra gauge.

Section two is reserved for the construction of the lagrangian
explicitly in terms of the solvable Lie algebra gauge scalar
fields. In section three we derive the field equations. We also
shortly inspect the algebraic properties of certain matrix terms
which appear in the field equations.
\section{Lagrangian in the Solvable Lie Algebra Parameterization}\label{section1}
The symmetric space sigma model (SSSM) is based on the coset
manifold $G/K$ where $G$ is in general a non-compact real form of
any other semi-simple Lie group and $K$ is a maximal compact
subgroup of it \footnote{We will consider the left cosets.}. The
coset manifold $G/K$ is a Riemannian globally symmetric space for
all the $G$-invariant Riemannian structures on it \cite{hel}. In
this section we will present the construction of the SSSM
lagrangian in the current formalism under the solvable Lie algebra
gauge \cite{fre}. Now consider the set of $G$-valued maps
$\nu(x)$. We assume that they transform as $\nu\rightarrow k(x)\nu
g$, $\forall g\in G$, $k(x)\in K$. The maps $\nu(x)$ are from the
$D$-dimensional spacetime into the group $G$. We further assume
that the maps $\nu(x)$ correspond to a parametrization of the
coset $G/K$. Thus their images are the representatives of the left
cosets of $G/K$. In the solvable Lie algebra gauge \cite{fre} the
parametrization $\nu(x)$ of the coset $G/K$ can be chosen as
\begin{equation}\label{ss1}
\nu(x)=e^{\varphi^{i}(x)T_{i}},
\end{equation}
where the basis $\{T_{i}\}$ generates a solvable Lie subalgebra
$\mathbf{s}$ of $\mathbf{g}$ which is the Lie algebra of $G$ and
$\{\varphi^{i}(x)\}$ are scalar fields on the $D$-dimensional
spacetime. The solvable Lie algebra $\mathbf{s}$ emerges from the
Iwasawa decomposition of $\mathbf{g}$ which reads \cite{hel}
\begin{equation}\label{ss2}
\mathbf{g}=\mathbf{k}\oplus \mathbf{s},
\end{equation}
where $\mathbf{k}$ is the Lie algebra of $K$. Explicit basis
constructions for $\mathbf{s}$ within the root space decomposition
of $\mathbf{g}$ can be referred in \cite{ker2,nej2,hel}. In this
work we will consider a generic basis $\{T_{i}\}$ for an arbitrary
Iwasawa decomposition \eqref{ss2}. One of the consequences of the
Iwasawa decomposition \eqref{ss2} is the local diffeomorphism
\cite{hel}
\begin{equation}\label{ss3}
Exp: \mathbf{s} \longrightarrow G/K,
\end{equation}
from the ${\Bbb{R}}^{dim\mathbf{s}}$-manifold $\mathbf{s}$ into
$G/K$ which enables the definition of the coset map \eqref{ss1}.
In the vielbein formalism of the SSSM the construction of the
lagrangian which is invariant under the right rigid action of $G$
and the left local action of $K$ is based on the introduction of
the Cartan-form
\begin{equation}\label{ss4}
{\mathcal{G}^{\prime}}=\nu^{\prime -1}d\nu ^{\prime}=P+Q,
\end{equation}
where
\begin{equation}\label{ss5}
P=P^{i}F_{i}\quad,\quad Q=Q^{j}K_{j}.
\end{equation}
$\{K_{j}\}$ is a basis for $\mathbf{k}$ and $\{F_{i}\}$ is the
basis which generates the orthogonal complement of $\mathbf{k}$ in
a vector space direct sum of $\mathbf{g}$. The fields $\{P^{i}\}$
form a vielbein of the $G$-invariant Riemannian structures on
$G/K$ and $\{Q^{j}\}$ can be considered as the connection
one-forms of the gauge theory over the $K$-bundle. Further
transformation properties of $P$ and $Q$ can be referred in
\cite{westsugra,nej2}. An invariant lagrangian under the above
mentioned global and local transformations can be constructed as
\cite{westsugra,tani}
\begin{equation}\label{ss6}
{\mathcal{L}}=\frac{1}{2}\, tr(\ast P\wedge P).
\end{equation}
Now we should remark an important aspect of the solvable Lie
algebra gauge. In the general parametrization of the coset the
coset generators need not form a subalgebra \cite{westsugra,tani}.
For this reason in general the Cartan-form \eqref{ss4} has
components both in $P$ and $Q$ directions since the commutation of
the coset generators may have components along the basis
$\{K_{j}\}$. On the other hand when we take the basis $\{T_{i}\}$
which generates the solvable Lie algebra $\mathbf{s}$ to be the
coset generators \footnote{This will be our choice in this work.}
then as the basis elements $\{T_{i}\}$ close on themselves we have
a major simplification in the calculation of the Cartan-form
\eqref{ss4}.

On the other hand one can also construct the symmetric space sigma
model in the internal metric formalism
\cite{julia1,julia2,ker1,ker2,nej1,nej2}. In \cite{sssm1} the
field equations of the internal metric formalism of the symmetric
space sigma model are derived for the axion-dilaton
parametrization. In this work bearing in mind the above discussion
of the simplification of the calculation of the Cartan-form we
will consider the lagrangian
\begin{equation}\label{ss8}
{\mathcal{L}}=\frac{1}{2}\, tr(\ast \mathcal{G}^{\prime}\wedge
\mathcal{G}^{\prime}).
\end{equation}
This lagrangian that is based on the Cartan form \eqref{ss4} which
is a Noether's current\footnote{Although it can be expressed in
terms of the Cartan-form since its kinetic term is directly
written via an internal metric we use the name of internal metric
formalism for the lagrangian studied in \cite{sssm1}. On the other
hand we call the lagrangian \eqref{ss8} to be the current
formalism since it is directly constructed from the Cartan-form.}
can also be obtained from the one in \cite{sssm1} by local field
redefinitions following the discussions of \cite{nej2} however on
its own right it formulates the theory for a generic solvable Lie
algebra parametrization in the form \eqref{ss1}. As we mentioned
before the Cartan-form \eqref{ss4} is already calculated in
\cite{nej2} it reads
\begin{equation}\label{ss9}
\mathcal{G}^{\prime}=\overset{\rightharpoonup }{
\mathbf{T}}\:\mathbf{W}\:\overset{\rightharpoonup }{\mathbf{
d\varphi}},
\end{equation}
where the dim$\mathbf{s}\times$dim$\mathbf{s}$ matrix $\mathbf{W}$
is
\begin{equation}\label{ss10}
\mathbf{W}=(I-e^{-M})M^{-1}.
\end{equation}
In \eqref{ss10} we define the matrix $M$ as
\begin{equation}\label{ss11}
M_{\alpha}^{\beta}=\varphi^{i}C_{i\alpha}^{\beta}.
\end{equation}
Also the components of the row vector $\overset{\rightharpoonup }{
\mathbf{T}}$ are $T_{i}$ and $\overset{\rightharpoonup
}{\mathbf{d\varphi}}$ is a column vector of the field strengths
$\{d\varphi^{i}\}$. We introduce the structure constants of
$\mathbf{s}$ as
\begin{equation}\label{ss12}
[T_{i},T_{j}]=C_{ij}^{k}T_{k}.
\end{equation}
Now that we have the exact form of the Cartan-form at hand we can
explicitly write down the lagrangian \eqref{ss8} in terms of the
coset parameterizing scalars $\{\varphi^{i}\}$. Inserting
\eqref{ss9} in \eqref{ss8} yields
\begin{equation}\label{ss13}
\mathcal{L}=\frac{1}{2}T_{np}W^{n}_{l}\ast d\varphi^{l}\wedge
W^{p}_{k}d\varphi^{k},
\end{equation}
where without specifying any trace convention we have introduced
the generic trace convention coefficients
\begin{equation}\label{ss14}
T_{np}=tr(T_{n}T_{p}).
\end{equation}
\section{The Field Equations}\label{section2}
In this section starting from the lagrangian \eqref{ss13} we will
derive the field equations of the coset scalars $\{\varphi^{i}\}$
by direct variation. Beforehand we should examine the variation
properties of the matrix $\mathbf{W}$. Firstly we observe that
\begin{equation}\label{ss15}
 M^{\prime}\equiv\frac{\partial M}{\partial\varphi^{m}}=C_{m},
\end{equation}
where
\begin{equation}\label{ss16}
(C_{m})^{k}_{j}=C^{k}_{mj},
\end{equation}
is the matrix representative of the generator $T_{m}$ in the
adjoint representation of $\mathbf{s}$ that is induced by the
basis $\{T_{i}\}$. Now the variation of $e^{-M}$ with respect to
the field $\varphi^{m}$ yields \cite{sat,hall}
\begin{equation}\label{ss17}
\begin{aligned}
\frac{\partial e^{-M}}{\partial\varphi^{m}}&=-e^{-M}(\frac{e^{ad_{M}}-I}{ad_{M}})(M^{\prime})\\
\\
&=-e^{-M}(M^{\prime}+\frac{1}{2!}[M,M^{\prime}]+\frac{1}{3!}[M,[M,
M^{\prime}]]+\cdot\cdot\cdot).
\end{aligned}
\end{equation}
Before going further we should inspect the structure of the
commutation series in \eqref{ss17}. For this purpose we will take
a closer look at the properties of the solvable Lie algebra
$\mathbf{s}$. The set
\begin{equation}\label{ss18}
{\mathcal{D}}\mathbf{s}=\{[X,Y]\},
\end{equation}
which is generated by all the elements $X,Y\in \mathbf{s}$ is an
ideal of $\mathbf{s}$. Since an ideal is a subalgebra
${\mathcal{D}}\mathbf{s}=[\mathbf{s},\mathbf{s}]$ is called the
derived algebra of $\mathbf{s}$. The higher order derived algebras
are defined inductively in the same way over one less rank derived
algebra and they are denoted as ${\mathcal{D}}^{n}\mathbf{s}$
where
\begin{equation}\label{ss19}
{\mathcal{D}}^{n}\mathbf{s}={\mathcal{D}}({\mathcal{D}}^{n-1}\mathbf{s}),
\end{equation}
 and ${\mathcal{D}}^{0}\mathbf{s}=\mathbf{s}$. Since the Lie algebra $\mathbf{s}$ is solvable
there exists an integer $n\geq 0$ such that
${\mathcal{D}}^{n}\,\mathbf{s}=\{0\}$. As we have mentioned before
one can construct a basis for $\mathbf{s}$ by using the root space
decomposition of $\mathbf{g}$. From \cite{ker2,nej2} we have
\begin{equation}\label{ss20}
\mathbf{s}=\mathbf{h_{p}}\oplus \mathbf{n},
\end{equation}
where $\mathbf{h_{p}}$ is a subalgebra of the Cartan subalgebra of
$\mathbf{g}$ and $\mathbf{n}$ is a nilpotent subalgebra which is
generated by certain positive root generators. For the explicit
construction of this decomposition we refer the reader to
\cite{ker2,nej2,hel}. For our purposes in this work it suffices to
observe that
\begin{equation}\label{ss21}
[\mathbf{h_{p}},\mathbf{h_{p}}]=0,\quad
[\mathbf{n},\mathbf{n}]\subset\mathbf{n},\quad[\mathbf{h_{p}},\mathbf{n}]=\mathbf{n},
\end{equation}
from which we conclude that the first derived algebra of
$\mathbf{s}$ is the nilpotent algebra $\mathbf{n}$ namely
\begin{equation}\label{ss22}
{\mathcal{D}}\mathbf{s}=[\mathbf{s},\mathbf{s}]=\mathbf{n}.
\end{equation}
On the other hand the image of the nilpotent Lie subalgebra
$\mathbf{n}$ in the adjoint representation $ad(\mathbf{n})$ is
also nilpotent \cite{hel} thus the central descending series
\begin{equation}\label{ss23}
\varphi^{0}ad(\mathbf{n})\supset\varphi^{1}ad(\mathbf{n})\supset\varphi^{2}ad(\mathbf{n})\supset\cdot\cdot\cdot,
\end{equation}
terminates with $\varphi^{m}ad(\mathbf{n})=\{0\}$ for some $m\geq$
dim$(ad(\mathbf{n}))$ \cite{hel,carter,onis}. In \eqref{ss23} the
ideals are defined as
\begin{equation}\label{ss24}
\varphi^{p+1}ad(\mathbf{n})=[ad(\mathbf{n}),\varphi^{p}ad(\mathbf{n})],
\end{equation}
with $\varphi^{0}ad(\mathbf{n})=ad(\mathbf{n})$. From their
definitions in \eqref{ss11} and \eqref{ss15} we deduce that $M$
and $M^{\prime}$ lie in the adjoint representation of $\mathbf{s}$
that is induced by $\{T_{i}\}$. From \eqref{ss22} we have
\begin{equation}\label{ss25}
\mathcal{D}ad(\mathbf{s})=[ad(\mathbf{s}),ad(\mathbf{s})]=ad(\mathbf{n}).
\end{equation}
Now also bearing in mind that the derived algebra
${\mathcal{D}}ad(\mathbf{s})$ is an ideal we have
\begin{equation}\label{ss25.5}
[ad(\mathbf{s}),\mathcal{D}ad(\mathbf{s})]=
\mathcal{D}ad(\mathbf{s})=ad(\mathbf{n}).
\end{equation}
Therefore we conclude that except the first one the rest of the
terms in the series \eqref{ss17} all lie in
${\mathcal{D}}ad(\mathbf{s})=ad(\mathbf{n})$. There always exists
a basis $\{T_{j}\}$ which induces a matrix representation such
that the matrix representatives of the nilpotent endomorphisms
which form up the nilpotent algebra $ad(\mathbf{n})$ have null
entries on and below the diagonal. Such a basis choice will
contribute a major simplification to the calculation of
\eqref{ss17}. Even for a matrix representation induced by a
generic basis the commutation terms in \eqref{ss17} will produce
images which belong to a fixed generation of matrices due to the
closure relations in \eqref{ss21} and the terminating central
descending series structure in \eqref{ss23} of $ad(\mathbf{n})$.
Thus in general although the series in \eqref{ss17} does not
terminate after a finite number of terms the calculation of the
series \eqref{ss17} will reduce to the calculation of the
coefficient series of the periodically appearing matrices in the
above mentioned set of finite number of generations. Now we are
ready to vary the matrix function $\mathbf{W}$. After some algebra
we find
\begin{equation}\label{ss26}
\begin{aligned}
\mathcal{K}_{m}\equiv\frac{\partial\mathbf{W}}{\partial\varphi^{m}}\:=&\:
e^{-M}(\frac{e^{ad_{M}}-I}{ad_{M}})(M^{\prime})M^{-1}\\
\\
&-\mathbf{W}M^{\prime}M^{-1},
\end{aligned}
\end{equation}
where we have made use of the identity
\begin{equation}\label{ss27}
\frac{\partial
M^{-1}}{\partial\varphi^{m}}=-M^{-1}M^{\prime}M^{-1}.
\end{equation}
Equipped with the above variation tools we can now directly vary
the lagrangian \eqref{ss13} with respect to the scalar field
$\varphi^{m}$. The variation yields
\begin{equation}\label{ss28}
\begin{aligned}
(-1)^{(D-1)}&d(T_{np}(\mathbf{W}^{n}_{k}\mathbf{W}^{p}_{m}+
\mathbf{W}^{n}_{m}\mathbf{W}^{p}_{k})\ast d\varphi
^{k})\\
\\
&=T_{np}(\mathcal{K}^{n}_{ml}\mathbf{W}^{p}_{k}+
\mathbf{W}^{n}_{l}\mathcal{K}^{p}_{mk})\ast d\varphi^{l}\wedge
d\varphi^{k}.
\end{aligned}
\end{equation}
These are the field equations for the solvable Lie algebra gauge
scalars $\{\varphi^{i}\}$ of the current formalism of the
symmetric space sigma model. We observe that the non-linearity of
the theory is highly reflected in the matrix component
coefficients in \eqref{ss28}.
\section{Conclusion}
By adopting the solvable Lie algebra gauge to parameterize the
coset manifold $G/K$ we have explicitly constructed the lagrangian
of the symmetric space sigma model in the current formalism in
terms of the coset scalar fields. The formulation makes use of the
exact form of the Cartan-form whose components are derived as
functions of the coset parameterizing scalar fields in
\cite{nej2}. Having expressed the lagrangian in terms of the
scalar fields we have varied it to obtain the corresponding field
equations.

The scalar fields in the supergravity theories are either elements
of the building block multiplets or they arise as components of
the dimensional reduction ansatz in the compactification scheme of
higher dimensional theories. Therefore they have a more
fundamental role than the vielbein. This fact justifies the
importance of deriving the field equations explicitly in terms of
the scalars. Our formulation is performed for a general symmetric
space $G/K$ and it is purely in terms of the unspecified solvable
Lie algebra structure constants. We have not also assumed a
specific representation. Thus we have constructed a general
formalism to derive the field equations of the symmetric space
sigma model in the Noether's current approach. Obtaining the field
equations is a systematic task. For a specific example one only
needs to identify the solvable Lie algebra which takes part in an
Iwasawa decomposition of the global symmetry algebra and then to
specify a basis for the solvable Lie algebra which gives the
structure constants that are the keys of the formulation.

We observe that due to the matrix structure of the field equations
there is a high degree of non-linearity. However one may express
the field equations as a matrix equation and one may make use of
the adjoint representation and perform certain field
transformations to simplify the coefficients of the field
strengths and the kinetic terms of the scalars which may abolish a
degree of coupling in the equations. One may also work on the
first-order formulation of the theory. An axion-dilaton
parametrization of the coset manifold in the current formalism can
separately be studied.

\end{document}